\documentclass[fleqn,usenatbib]{mnras}

\usepackage{newtxtext,newtxmath}
\usepackage[T1]{fontenc}
\usepackage{ae,aecompl}

\usepackage{graphicx}	
\usepackage{amsmath}	
\usepackage{amssymb}	
\usepackage{multirow}

\newcommand{\dmunits}{{\rm pc\,cm^{-3}}}
\usepackage{color}

\newcommand{\eg}{e.\,g.}

\title[FETCH: A DL based fast transient classifier]{FETCH: A deep-learning based classifier for fast transient classification}
\author[Agarwal et al. (2019)]{
Devansh Agarwal,$^{1,2}$\thanks{Both authors contributed equally to this work.} \thanks{E-mail: da0017@mix.wvu.edu (DA)}
Kshitij Aggarwal,$^{1,2}$\footnotemark[1]\thanks{E-mail: ka0064@mix.wvu.edu (KA)}
Sarah Burke-Spolaor$^{1,2}$,
\newauthor{Duncan R. Lorimer$^{1,2}$ and Nathaniel Garver-Daniels$^{1,2}$}
\\
\footnotetext{Both authors contributed equally to this work.}
$^{1}$West Virginia University, Department of Physics and Astronomy, P. O. Box 6315, Morgantown 26506, WV, USA\\
$^{2}$Center for Gravitational Waves and Cosmology, West Virginia University, Chestnut Ridge Research Building, Morgantown 26506, WV, USA\\
}

\date{Accepted 2020 June 23. Received 2020 June 22; in original form 2019 March 3}

\pubyear{2019}

\begin{document}
\label{firstpage}
\pagerange{\pageref{firstpage}--\pageref{lastpage}}
\maketitle
\begin{abstract}
With the upcoming commensal surveys for Fast Radio Bursts (FRBs), and their high candidate rate, usage of machine learning algorithms for candidate classification is a necessity. Such algorithms will also play a pivotal role in sending real-time triggers for prompt follow-ups with other instruments. In this paper, we have used the technique of Transfer Learning to train the state-of-the-art deep neural networks for classification of FRB and Radio Frequency Interference (RFI) candidates. These are convolutional neural networks which work on radio frequency-time and dispersion measure-time images as the inputs. We trained these networks using simulated FRBs and real RFI candidates from telescopes at the Green Bank Observatory. We present 11 deep learning models, each with an accuracy and recall above 99.5\% on our test dataset comprising of real RFI and pulsar candidates. As we demonstrate, these algorithms are telescope and frequency agnostic and are able to detect all FRBs with signal-to-noise ratios above 10 in ASKAP and Parkes data. We also provide an open-source python package \texttt{FETCH} (Fast Extragalactic Transient Candidate Hunter) for classification of candidates, using our models. Using \texttt{FETCH}, these models can be deployed along with any commensal search pipeline for real-time candidate classification.
\end{abstract}




\section{Introduction}

Fast Radio Bursts (FRBs) are extremely bright, millisecond-duration radio transients that are characterised by dispersion measures (DMs) that are much higher than the expected Milky Way contribution originally seen in data from the Parkes radio telescope \citep{lorimer07,thronton2013}. They have subsequently been detected in data collected at Arecibo \citep{spitler14},  Green Bank Telescope (GBT) \citep{masui15}, the upgraded Molonglo Synthesis Telescope (UTMOST) \citep{caleb17}, and the Australian Square Kilometre Array Pathfinder (ASKAP) \citep{bannister17, shannon18}. Of over 60 FRBs published,\footnote{http://frbcat.org \citep{petroff16}} two have been found to repeat: FRB~121102 \citep{spitler16} and FRB~180814.J0422+73 \citep{r2amiri2019}. FRB~121102 was confidently localized to a  low-metallicity host galaxy at a redshift of 0.19 by the {\sc Realfast} detector \citep{law2018} on the Karl G.~Jansky Very Large Array \citep{chatterjee17, tendulkar17}, making it evident that some, if not all, FRBs are cosmological in origin.

FRB searches are typically done on high time and frequency resolution radio astronomical data by first correctly accounting for the dispersive delay over many trial DM values. This is then frequency averaged to generate a time series. These de-dispersed time series are then convolved with box-car kernels of various widths to look for broader pulses. Finally, candidates above a detection threshold are marked for visual inspection by a human. More recently, however, with the advent of state-of-the-art de-dispersion algorithms and Graphic Processing Unit (GPU)-accelerated pipelines (e.g., {\sc heimdall}\footnote{https://sourceforge.net/projects/heimdall-astro} \citep{barsdell2012}; FREDDA (Bannister et al. in prep); bonsai (Smith et al. in prep)), it is now possible to implement real-time FRB searches. As a result, commensal back-ends for FRB detection are now running on many radio telescopes around the world.
All of these searches are affected by a high false positive rate, due both to Gaussian noise and the presence of Radio Frequency Interference (RFI), which can generate up to thousands of candidates per day. Hence, manual inspection of all FRB candidate becomes challenging and infeasible. To reduce the sheer volume of candidates that require inspection, a number of techniques are presently being applied. These often include basic RFI mitigation techniques to remove ${\rm DM} = 0\,\dmunits$ signals or other common types of RFI \citep{eatough2009,nita2010, dumez2016}. Clustering algorithms like k-Nearest neighbours \citep{cover2006} and friends-of-friends \citep{ester1996} have also long been deployed to identify single, bright events that trigger many candidates \citep[see, e.g.,][]{deneva+09,burkespolaor+11}. 

However, the above-stated techniques cannot \emph{classify} candidates, for example as RFI, FRB or pulsars. Traditionally, classification of candidates has been done manually, which limits the ability to trigger real-time multi-wavelength follow-ups, and forces a requirement to record and store large data volumes. Machine learning has the potential to provide an automated solution to this problem. Moreover, machine learning techniques have already been widely used for signal classification and pattern recognition. Deep learning is a part of machine learning methods based on learning data representations, as opposed to task-specific algorithms.
Deep learning has already been applied to pulsar searches \citep{zhu14, guo17, devine16, bethapudi18, mcfadden18} yielding significant improvements, demonstrating their potential for use in transient searches. \citet{wagstaff16} and \citet{foster18} have applied a supervised random forest classifier by extracting data-specific features to classify the candidates into certain pre-defined classes of RFI and FRBs. Recently, \citet{zhang18} and \citet{connor18} have used convolutional neural networks for FRB classification. 


In this paper, we present a set of deep neural networks developed using the approach of transfer learning. We have utilised the state-of-the-art models trained for real-world object recognition in images to classify single pulses (eg: FRBs and pulsars) and RFI in fast-transient search data. In this work, single pulses from FRBs and pulsars are considered alike, and the models do not differentiate between the two. FRBs and pulsar single pulses can be differentiated in post processing based on their detection DM. If the DM of the pulse is greater than the Galactic DM in that line of sight, then the pulse could be of extragalactic origin (i.e. a FRB) else from a pulsar. Our networks use frequency-time and DM-time images as inputs. These networks are telescope and frequency agnostic in nature and can classify candidates in real time. We provide an open source package \texttt{FETCH}, which can easily be integrated into any FRB search pipeline with minimal effort. The rest of this paper is organised in the following manner. In \S2 we provide a brief introduction to convolutional neural networks and transfer learning. In \S3 we detail the  data used for training and testing the algorithms and in \S4 describe the methods. Results are detailed in \S5, followed by a discussion in \S6 and a roundup of our main conclusions in \S7.

\section{Machine Learning}

Machine learning gives computer systems the ability to ``learn'' from data, without being explicitly programmed using statistical techniques. Artificial neural networks are a class of models within the general machine-learning framework, which is itself based on biological neural networks. They have revolutionized machine learning. Here, we provide a brief introduction to Convolutional Neural Networks (CNN) and refer the reader to a more detailed description by \citet{goodfellow16}.

\subsection{Neural Networks}

As mentioned above, neural networks are a type of machine learning algorithms, inspired by the biological neuron, and their network. A neural network typically consists of many different types of layers. The number of layers correspond to the \textit{depth} of the network. Each layer is made up of many `neurons' connected to the output of previous layer. The neuron is the fundamental unit of the network. Each neuron performs a weighted sum of its inputs ($\mathbf{x}$) and returns an output 

\begin{equation}
\mathbf{y} = f(\mathbf{w}\cdot \mathbf{x}+{b}),
\end{equation}

where $\mathbf{w}$ represents the weights used, the function $f$ is a non-linear mathematical operation referred to as an ``activation function/layer'' which decides the output of that layer and $b$ is the bias. Both weights and bias are \textit{learned} during training. There are many types of activation functions. One example is 
\begin{equation}
f(x) = \max (0,x)
\end{equation}
which is usually referred to as the Rectified Liner Unit (ReLU).
The types of layers and their arrangement makes up the architecture of the neural network. For example, a layer in which all the inputs are connected to all the outputs, is called a \textit{dense} layer. 
A combination of weights and bias along with a multi-layered network could be used to map the given inputs to the known outputs (or labels) using any non-linear function.   
\subsection{Convolutional Neural Networks}
\begin{figure}
	\includegraphics[width=84mm]{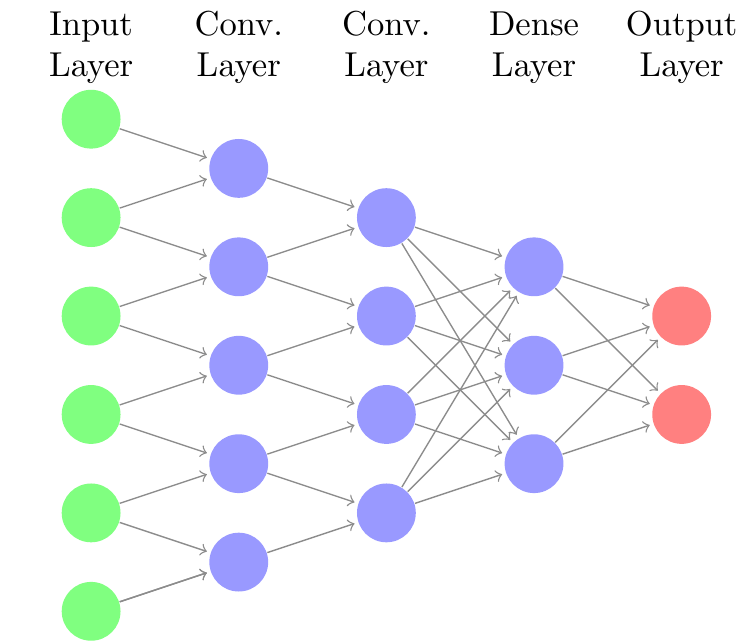}
    \caption{Simplified schematic representation of a CNN architecture. Each circle represents a neuron in the network. The arrows depict the connections between the neurons. Two convolutional layers are labelled as ``Conv. Layer''. These are followed by a dense layer and an output layer (see text for details).}
    \label{fig:cnn}
\end{figure}
\label{sec:cnn}
CNNs are a type of artificial neural network used for working with images. A fundamental issue with using conventional neural networks on images is that number of independent weights required to connect two consecutive layers increases exponentially with input image sizes. A CNN on the other hand, uses convolutional layers, 
which are a set of many small kernels (filters) that are convolved with images. This is then followed by an ``activation function", as described in the previous subsection. 
The convolution and the activation layer together extract features from the inputs. These are then fed into a pooling layer where the image size is reduced, by either averaging or taking the maximum of a few adjacent pixels. More such sets of convolution, activation and pooling layers are applied. Finally, the processed images are reshaped to a one-dimensional array where they are connected to a dense layer. This is where the extracted features 
are assigned likelihoods of belonging to distinct output classes. Fig.~\ref{fig:cnn} shows a simplified version of a CNN. The network shows two convolutional layers and two dense layers. 
Typically, the end-result of this process is a probability of the input candidate belonging to various classes. The network structure is determined by the total number of layers, number of convolutional filters, the number of units in the dense layer and the choice of activation function. These are called \textit{hyperparameters}, their choice is dependent on specific application. 

For a CNN to make useful predictions, we first train them using labelled data. The initial weights for all the filters and dense layers are set to be random numbers. The labelled data are then passed through the network, and the classification probabilities are obtained. This is called \textit{forward propagation}. 
The deviation of a true label with respect to the probability given by the network, is quantified using a so-called cost function. The cost function quantifies the error between predicted values and expected values and presents it in the form of a single real number. The cost function, when evaluated over the labelled data is, referred to as the \textit{loss}. To train the network, the loss is minimized with the help of an optimisation algorithm. Such algorithms compute the gradient of the cost function with respect to different weights. The gradient, coupled with a constant called the learning rate, is used to modify the weights. This is called \textit{backward propagation}. 

The labelled data are divided into three sets: training, validation and test data. The networks are trained as described above on the training data, and are evaluated based on its performance on the validation data. This is repeated until its performance on validation data are satisfactory. Once the network is trained, its performance is reported on the test data. Typically, these datasets consist of a few thousand examples. Note that the validation and test datasets are never used to train the network. 

Both forward and backward propagation are computationally intensive processes, and hence are done in small batches. When the complete training data is passed once through forward and backward propagation, its called an \textit{epoch}. Training CNN is an iterative process requiring several such epochs and is often done on GPUs. The training process is stopped when the desired performance criterion is met (e.g.~$\sim 99$\% accuracy). While training, one often runs into one of the three cases: underfit, overfit or robust fit. Underfitting occurs when performance on both training and validation data is poor. This usually implies that the model needs to have more parameters to fit the data. When training performance is outstanding and the performance on validation data is poor, it falls under the overfitting regime. In this case, rather than learning to recognise the general pattern, the model has memorised the specific patterns of training data. This is often the case with neural networks, as they intrinsically have a large number of parameters. This can be solved by reducing the network size, getting more training data, or by penalising the network heavily for an incorrect classification. This is called regularisation \citep{krogh1991}. Lastly, when the training and validation performance are similar, it is said to be a robust fit. 

Once the model is trained, forward propagation is used to obtain classification probabilities for a given input. This is called \textit{inference}.   
\subsection{Why deeper is better?}

Both FRBs and RFI exhibit complex structure in frequency-time and DM-time space. A deeper network has more layers in it and can learn features at various levels of abstractions \citep{mhaskar2016}. Multiple layers are better at generalising as they learn all the features starting from simple features in raw data, to high-level classification. 
Also, it has been shown that for a fixed number of parameters, going deeper allows the model to capture richer features \citep{eldan2015power}. However, this comes with a caveat: deeper models with a large number of trainable parameters are more likely to overfit if the input dataset is small. We address this problem using transfer learning. 

The ultimate goal of this analysis is to classify FRBs from RFI with the highest possible recall (see \S\ref{subsec:metrics}). Previously, both traditional machine learning and neural network-based classifiers have been used; however, they were specifically tuned for a particular telescope. In this paper, we show that by using deeper (in comparison to previously published) neural networks, we have created a telescope agnostic classifier which gives a high recall as well.

\begin{figure*}
	\includegraphics[width=168mm]{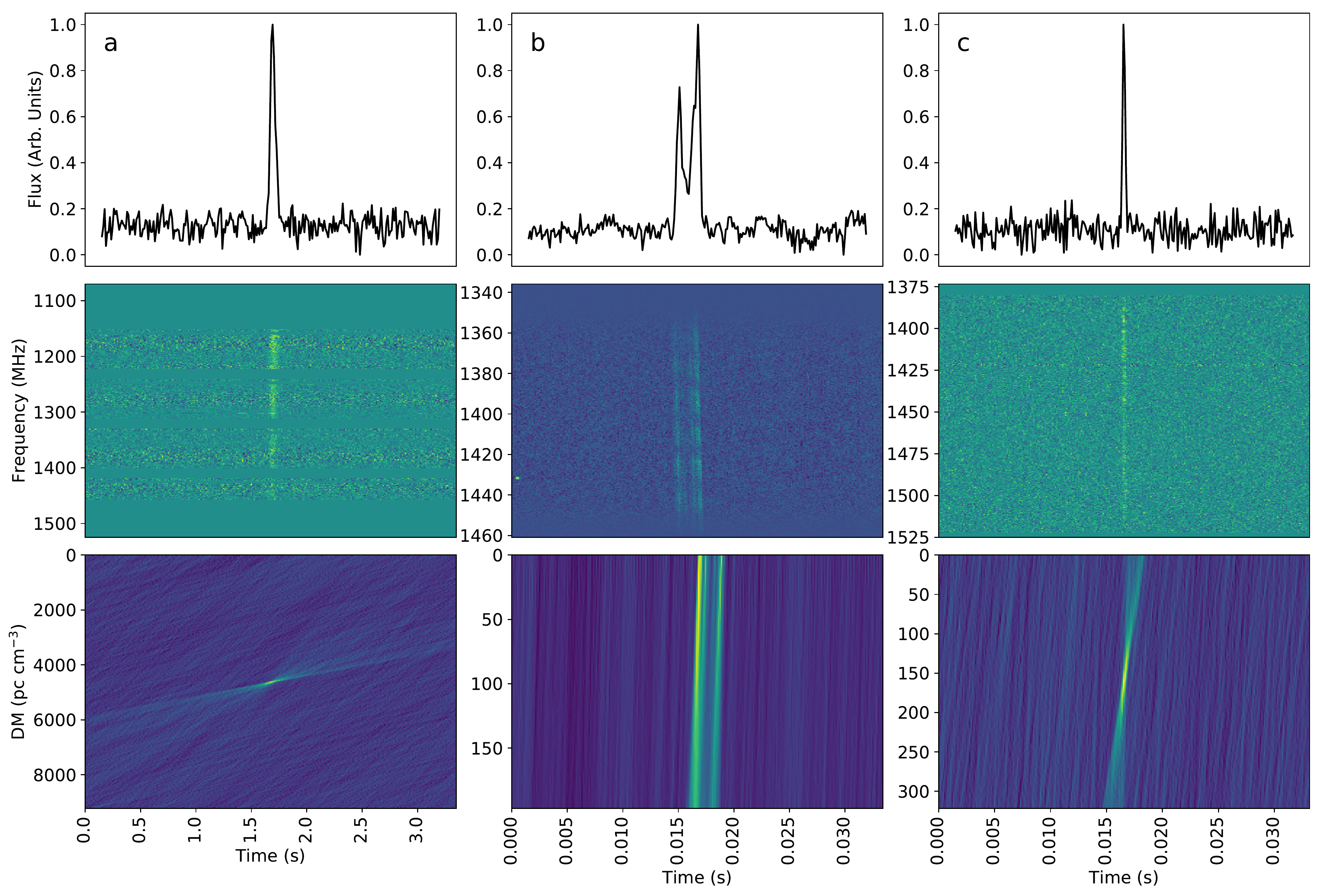}
    \caption{Sample images of high S/N candidates from the training and test dataset. The top row shows the time-series profile which is not included in our algorithms but is included for visual reference here. The middle row is the frequency-time image, while the bottom row is the DM-time image. Column (a) corresponds to a simulated FRB with background data from FLAG. The gaps in the frequency-time plots are due to instrumental effects. Column (b) is a real RFI candidate from the 20m telescope at the Green Bank Obesvatory. Column (c) is a pulsar observed using the FLAG system. Panels (a) and (c) represent the positive while panel (b) represents the negative examples in our case.}
    \label{fig:3_3}
\end{figure*}
\subsection{Transfer Learning}

The amount of data required to train a network is dependent on the number of trainable parameters of the model. Deep neural networks have tens of millions of parameters. Hence, the training sets typically consist of millions of data samples. More often than usual, such large datasets are not available for specific applications. Therefore, an alternative technique of transfer learning is employed. Here, pre-trained networks are used to extract features. In networks, the initial few layers learn to identify basic features like edges, the following layers would learn a collection of edges or shapes, while even deeper layers would train itself on the collection of shapes, and the subsequent layers will learn even higher-order features.

The classification takes place in the final dense layer. This layer is replaced by a custom dense layer with the number of units equal to classes of data. The convolutional layers remain frozen, i.e.~their weights will not change during training. The new classification layer can now be trained for the new dataset. Here, the model extracts features from the pre-trained convolutional layers and learns to map them to new classes in the dense layer. As we only need to train the final dense layer, the number of trainable parameters reduce significantly and a smaller dataset can be used for training. Also, depending on the size and features of the training dataset, one can ``fine tune'' the later convolutional layers as well. 

Transfer learning has been successfully used in various domains of astronomy, e.g.~identification of Supernovae Ia \citep{vilalta2018}, detecting galaxy mergers \citep{ackermann2018} and galaxy classification schemes \citep{aniyan2017, carrasco2018, khan2018}. Here we demonstrate the capability of transfer learning to develop a network for generically classifying FRB and non-FRB (RFI) events. We show that, by banking on the generic feature extraction of the pre-trained models and standardising the training dataset, the network becomes agnostic to the choice of both the observing frequency and the telescope/data acquisition device used. 



\subsection{Deep Learning for Transient Detection}
\label{subsec:DLinTD}
To help set the context of our work and how it extends previous efforts, we now describe the basic details of two previously published implementations of Deep Learning techniques for transient detection. 

\cite{connor18} have used a multi-input CNN with two convolution and two pool layers, with four inputs, namely:
\begin{itemize}
    \item de-dispersed frequency-time spectrograms (see Fig.~\ref{fig:3_3}) of a fixed size ($32 \times 64$ pixels); 
    \item DM-time grids that reflect the signal-to-noise ratio of the de-dispersed, frequency-averaged timeseries a function of DM and time (100 $\times$ 64 pixels); a DM-time plot is showed in the bottom row of Fig.~\ref{fig:3_3};
    \item the dedispersed, frequency-averaged timeseries itself (as in the top row of Fig.~\ref{fig:3_3}); 
    \item the signal-to-noise ratio as a function of position on the sky.
\end{itemize}
The authors demonstrated that their network worked well on the Apertif and CHIME telescopes after training separately for both.  
\cite{zhang18} developed a 17-layer \texttt{ResNet} architecture \citep{he15} using only the dispersed frequency-time spectrograms of size 352$\times$256 pixels as input. Using this model on the Breakthrough Listen C band observations of FRB~121102 \citep{gajjar2018}, they were able to find 93 bursts, of which 72 were new pulses. Both of the above networks perform excellently for their respective telescope back-ends. In this paper, we provide 11 generic models, which can be used at different radio telescopes irrespective of the observing frequencies. 



\section{Data used for this study}
\begin{table}
	\centering
	\caption{Instrument (backends), sources and number of candidates (including the augmented candidates) used for training, validating and testing both frequency-time (FT) and DM-time (DMT) inputs. T+V refers to the training and the validation data, Sim FRB stands for simulated FRBs.}
	\label{tab:datasets}
	\begin{tabular}{llccc} 
		\hline
		Instrument & Source & T+V & T+V & Test\\
		(back-end) &  & DMT & FT & \\
		\hline
        FLAG & & \\
        (FLAG) & RFI & 32,720 & 6,000 & 2,790 \\
        & Sim FRB & 20,000 & 8,500 & -\\
        & Pulsar & - & - & 2,288 \\
		\hline
        GBT L-Band & & \\
        (GREENBURST) & RFI & - & 6,000 & 2,170 \\
        & Sim FRB & 20,000 & 8,500 & - \\
        & Pulsar & - & - & 1,376 \\
        \hline
        Green Bank 20m & & \\
        (Skynet) & RFI & 9,854 & 8,000 & 2,359 \\
        (GBTrans) & Pulsar & - & 3,000 & 3,000 \\
        \hline
        Total & FRB & 40,000 & 20,000 & 6,664 \\
        & RFI & 42,574 & 20,000 & 7,319 \\
        \hline
	\end{tabular}
\end{table}
\subsection{Surveys}
\label{sec:surveys}
We used data from observations using Green Bank Telescope (GBT) and 20~m telescope both located at the Green Bank Observatory (GBO). The GBT data were recorded using commissioning test observations of GREENBURST \citep{surnis2019} and the pilot survey using the FLAG \citep{rajwade2019} instrument. The 20~m telescope data was observed using Skynet (\citet{Hosmer2013,Smith_2016};Gregg et al. in prep) and GBTrans \citep{Golpayegani2019} back-end. In order to create a uniform dataset we used {\sc heimdall} with the following parameters on all the above data: S/N $\geq$ 8, $10 < \rm DM < 10,000~\dmunits$ and width < 32~ms. It preforms a brute force dedispersion to transform data from frequency-time to DM-time space. Each dedispersed time series is baselined to zero mean and then sliding box-car filters of various widths are applied. The boxcar filtered time series is normalised to unit root mean squared deviation. Now, the peaks in the time series correspond to S/N, and a threshold is used to select the candidates. The generated candidates were manually labelled. From the above, we used 24,947 RFI candidates, 6000 Crab giant pulses from GBTrans, 1,931 and 357 pulses from B1933+16 and B2011+32, respectively, observed using FLAG. We also used 1,376 pulses from PSR B0740--28, detected with GREENBURST (see \S\ref{sec:test_train_data} for details).

While the above pulsar detections partly served as a training data set for astrophysical pulses, we also wished to train on signals that better represent FRBs: that is, typically isolated from other pulses in the data, and spanning a larger range in widths and DMs. Thus, to acquire a training data set that included such pulses, we injected simulated transients into around 2.4~h of data taken with GREENBURST (for MJD 58320) and 5.7~h from FLAG (between MJD 58146--58153). These data were selected randomly from various observations to ensure that they cover the broad variety of instrumental effects that typically impact observations. Examples of such effects are bandpass variations, nulling of part of the bandpass due to a malfunctioning subset of the telescope processing back-end, packet loss and low-level RFI.

\subsection{Simulating and Injecting FRBs}
\label{subsec:simulated_frb}
\begin{table}
	\centering
	\caption{Parameter Distribution for Simulated FRBs}
	\label{tab:sim_frb}
	\begin{tabular}{lcr} 
		\hline
		Parameter & Distribution & Range\\
		\hline
		Fluence (Jy~ms) & Log-normal  & $\mu=3.5$, $\sigma = 1$\\
        DM ($\dmunits$) & Uniform & 50, 5000 \\
        Width (ms) & Uniform &  0.5, 50\\
        Spectral Index & Uniform & -4, 4 \\
        Scattering Timescale & Uniform & 0, Width \\
		\hline
	\end{tabular}
\end{table}
\begin{figure}
	\includegraphics[width=84mm]{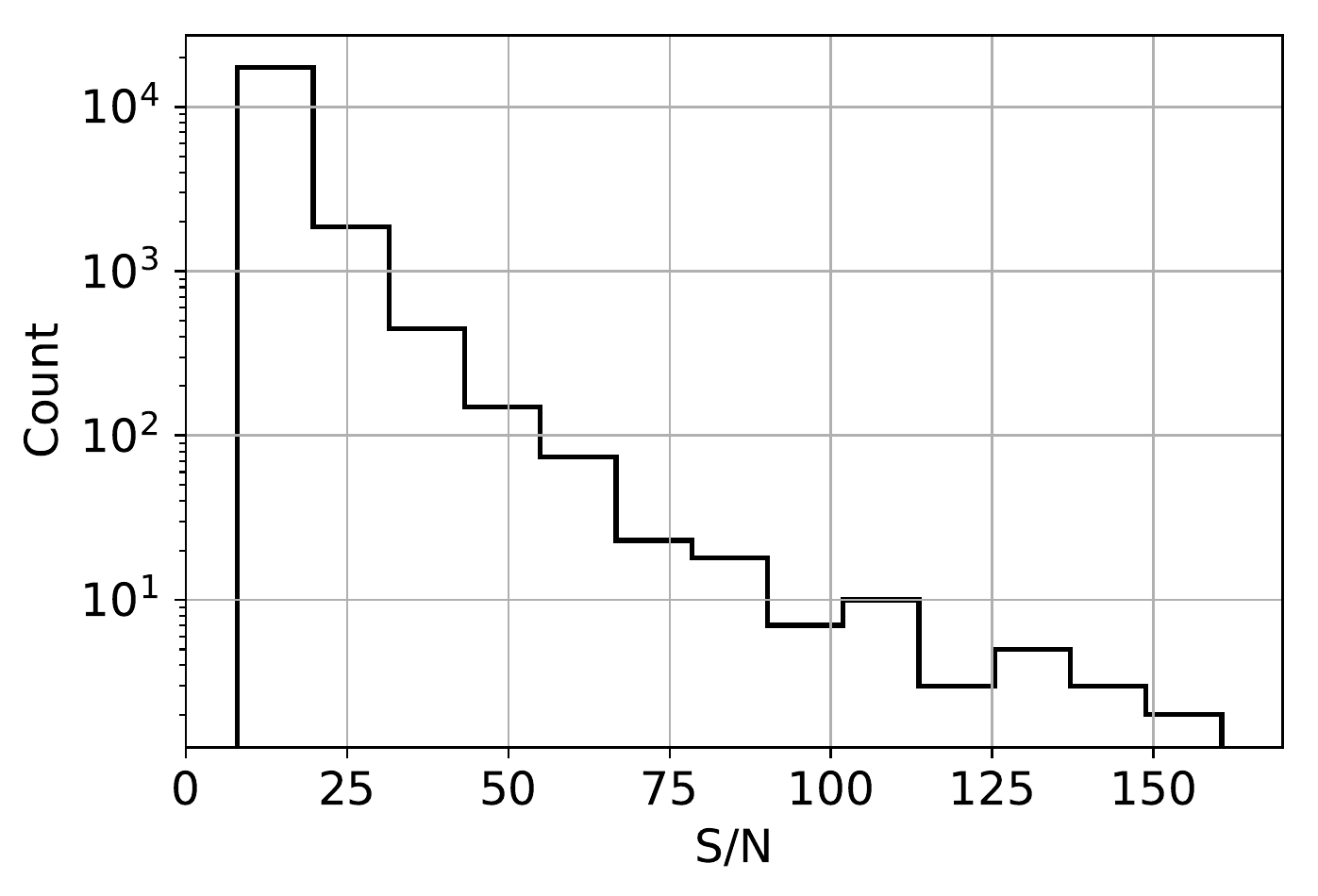}
    \caption{Distribution of S/N of the simulated FRBs.}
    \label{fig:injectedfrb}
\end{figure}
We chose the parameters of simulated FRB candidates from a predefined distribution (see Table \ref{tab:sim_frb}). Each pulse is then injected on randomly selected background data, as described above. After the injection, data were normalised to a median of zero and unit standard deviation. We then discard the candidates with an S/N less than 8. These codes to generate simulated FRBs were run on Super Computing System (Spruce Knob) at West Virginia University.
Fig.~\ref{fig:injectedfrb} represents the S/N distribution of the injected candidates after discarding the low-S/N events.


\subsection{Train and Test Datasets}
\label{sec:test_train_data}
Deep learning models, irrespective of their architecture, are heavily influenced by the size and quality of the dataset which is used to train them. For a binary classification application like ours (i.e.~``RFI'' vs.\ ``FRB''), it is advisable to have balanced training dataset, i.e.~nearly equal number of FRB and RFI candidates \citep{Buda2018}. Also, within each class, it is necessary to make sure that the features which are of interest (eg: vertical signal feature in the dedispersed frequency-time images, and bow-tie shape in DM-time images) are dominant in the images. We build upon the methods described in \citet{zhang18} and \citet{connor18} where the authors create a balanced train and test datasets using real RFI and simulated FRBs for training and testing their networks.


Table \ref{tab:datasets} provides the details of the candidates (including the augmented ones) in the datasets used for the Frequency-time (FT) and DM-time (DMT) models. We used 32,720 RFI candidates from FLAG and 9,854 RFI candidates from the Skynet backend towards the RFI examples to train the DMT models. For the same models, we used 20,000 simulated FRBs generated from GREENBURST and FLAG backends each. For training the FT models, we used 6000 RFI candidates from FLAG backend (randomly chosen from the original 32,720 RFI candidates), 6,000 RFI candidates from GREENBURST backend and 8,000 RFI candidates from Skynet backend. We used 8,500 simulated FRB candidates for both GREENBURST and FLAG backend each (randomly chosen from the original 20,000). We also used 3,000 giant pulses from Crab pulsar from GBTrans backend in this dataset used to train FT models. The test set was curated independently using separate observation scans with 2,790, 2170 and 2,359 RFI examples from FLAG, GREENBURST and Skynet backends respectively. Instead of using simulated FRBs, the test set contains pulsar single pulses which are listed as follows. From the FLAG backend, we used 357 and 1,931 single pulses from PSR~B2011+38 and PSR~1933+16 respectively. From the GREENBURST backend, we used 1,376 single pulses of PSR~B0740-28 and lastly, we used 3,000 Crab giant pulses from GBTrans backend.

The frequency-time images are dependent on the bandpass of individual back-ends, we balanced the number of candidates from each back-end as well. These variations can be seen from Fig. 4 in \citet{rajwade2019} and Fig. 2 in \citet{surnis2019}. In the DM-time images, as the frequencies are scrunched, the image is independent of such effects therefore we did not opt for any such balancing for it. On the other hand, FT images depend on the frequency structure of the data. Therefore, we made sure to use an equal number of RFI and simulated FRBs from GREENBURST and FLAG backends each. This was done to balance the features present in the individual backends. Due to this, the datasets for FT and DMT were not identical. We used the FT training dataset to train the combined models.

We split the training data randomly into 85\% training and 15\% validation sets. The random split of the data is justified because the data were taken from different backends on different days when the telescope was looking at different parts of the sky. These backends have different numbers of beams, bandwidth, observing frequency and time resolution. The spectrogram data in the images used for training the models are of the order of 10 seconds while the observations were spread across six months. As the two time scales differ by several orders of magnitude and the telescope was pointing at different locations with different backends at different times, it is highly unlikely that any two observations would be similar or correlated. The random shuffle of data makes sure that both the training and the validation data are drawn from the same distribution of features in the images.


The test dataset was used to evaluate and compare the performance of the combined models. It consists of real data, where we have used RFI and pulsars from different back-ends (see Table \ref{tab:datasets}). The dataset was sampled independently i.e the observation scans were different from those for training and validation data. Due to the limited number of pulsar candidates, we use a small number of those in training/validation set while keeping most of them for the test set. Furthermore, in \S\ref{subsec:eval_real} we detail a real-world test data set which includes data from the FRB searches from three telescopes: ASKAP, PARKES and GBT.

\subsection{Data Augmentation}
\label{sec:data_aug}
To expand on smaller data sets, and make the networks more robust, training data can be augmented in several ways to increase the number of candidates in your training data set. In the example of training images to recognize cats, one would expect a cat to be identified as such if it were facing rightward or leftward. Thus, the same image can be used twice in the training data (once as is and once inverted horizontally). 
Depending on the data and nature of the candidates (in particular its uniquely identifying features), this technique needs to be used with caution. For instance, one cannot typically horizontally invert FRB candidates because dispersion and scattering are not symmetric effects in time. 
However, we discuss here several aspects of this technique which can be applied to the radio transient data in the realm of RFI. We used both the techniques listed below to double the number of RFI candidates in each dataset. The number of candidates mentioned in Table \ref{tab:datasets} include augmented candidates.

\subsubsection{Frequency-Time Flip}
In de-dispersed data, the frequency-time image can be flipped along the time axis. This is because de-dispersion removes the dispersion asymmetry from the data. However, due to the presence of scattering, flipping along the frequency axis would not be advisable.

\subsubsection{DM-Time Flip}
DM-time data can be flipped along both time and DM axis. This would preserve the orientation of the bow-tie. Although, a DM-time flip is not physically meaningful, it is a useful technique from a computer vision point of view. 

\section{Methods}
\begin{figure*}
	\includegraphics[width=170mm]{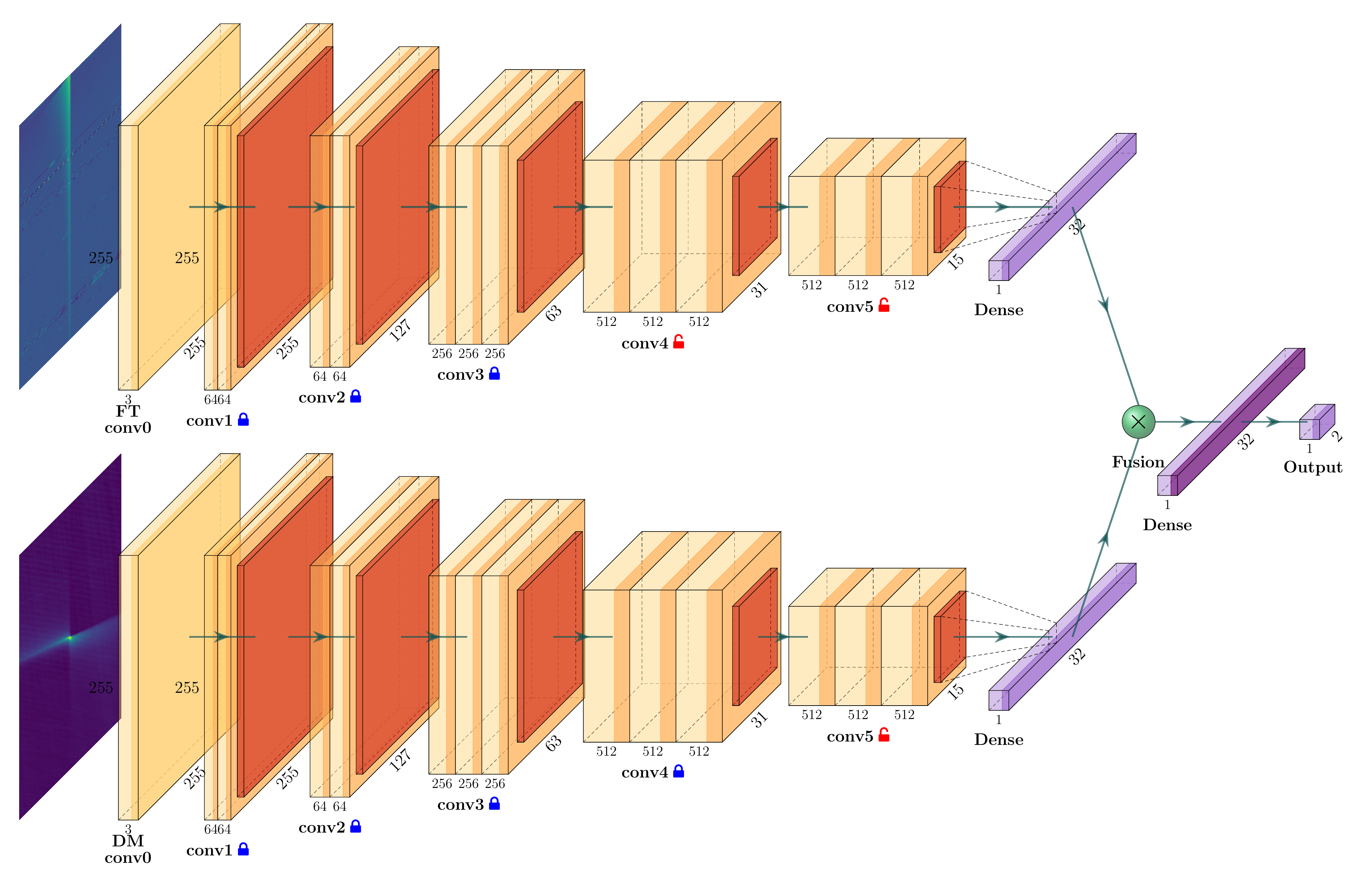}
    \caption{The figure shows a sample network architecture. The two inputs are the frequency-time and DM-time images. For simplicity, we have used the \texttt{VGG16} \citep{VGG} model to describe the architecture. The yellow boxes show the convolutional outputs and are labelled with output sizes. The brown edges represent the ReLU activation. The orange boxes depict the pooling layer. The dense layers are displayed in violet. The green ball represents the element-wise product of the two dense layers. The second last dense layer has a softmax activation function demonstrated by the darker coloured edge. The blue lock symbol represents the frozen layers while the red unlock symbol shows the unfrozen (i.e.~trainable) layers. The arrows show the network connections. The figure is generated using \citet{MakeCNN}.}
    \label{fig:arch}
\end{figure*}

In this section, we describe the network architectures, the data used for training and testing these networks, and a standardisation procedure. This ``standardisation'' refers to reshaping all input data to have the same size and shape. For instance, all spectrograms must have the same number of frequency channels and time samples to use in our trained algorithm. Following \citet{connor18}, we use frequency-time spectrograms and DM-time images as an input to our network. We train a different CNN for each input case and then combine the two (see \S4). In contrast, we do not use time-series data, as that information is already contained in frequency-time images. 
We have also opted not to use sky-dependent (\eg\ multi-beam) signal-to-noise as an input, because not all telescopes have this information available. Furthermore, we consider it a feasible alternative for sky-distributed RFI detections to be mitigated based on simple coincident rejection techniques, as multiple pipelines have done previously \citet{burkespolaor+11,champion2016,shannon18,amiri2019}.

\subsection{Input Data Standardization}
\label{sec:input_data_std}
We standardise our input data to make the algorithm agnostic to observing frequency and choice of the telescope. We use de-dispersed data in the frequency-time spectrogram as an input. Once de-dispersed, the data are independent of the original candidate DM and observational frequency (apart from any potential intrinsic frequency-dependent FRB properties, which may remain). We bin the time axis such that the candidate pulse profile lies between 1--4 bins of the origin. As a result, we are weakly sensitive to different sampling times on various telescope back-ends.
This also maximises the S/N by condensing the pulse to a few bins. The frequency-time image is then re-sized to 256$\times$256 pixels by averaging the frequency axis and trimming out the extra pixels. The choice of 256 frequency bins was made to preserve the frequency modulation of the recently reported FRBs \citep{shannon18, amiri2019, r2amiri2019, chatterjee17}. To reduce the effects of bandpass variation, we fit out a linear trend along the frequency axis.


DM-time images are created by scrunching (averaging along frequency axis) the frequency-time after de-dispersing it at different DMs. We chose the DM range from zero to twice the DM of the candidate, spread over 256 steps. The time axis was binned and cropped as explained above. A typical DM-time image of a real event looks like a bow-tie centered around a non-zero DM value. The edges of the bow-tie shape are bounded by the extent of the pulse profile. The angle between them is dependent on DM, the width of the candidate and the observing bandwidth. The area filled between these lines is governed by the spectra of the FRB. Fig.~\ref{fig:3_3} shows an example of the input images. 

\subsection{Network Architecture}
\label{sec:network_arch}
We use \texttt{keras} \citep{chollet2015keras} with the \texttt{TensorFlow}
\citep{tensorflow2015-whitepaper} back-end to develop our models for both frequency-time and DM-time inputs separately. \texttt{keras} provides the following networks with weights trained on Imagenet \cite{imagenet}. For consistency with the literature, we adopt the following acronyms:
\begin{itemize}
\item \texttt{Xception} \citep{Chollet16a}
\item \texttt{VGG16, VGG19} \citep{VGG}
\item \texttt{ResNet50} \citep{he15}
\item \texttt{DenseNet121, DenseNet169, DenseNet201} \citep{DenseNet}
\item \texttt{InceptionV3} \citep{InceptionV3}
\item \texttt{InceptionResNetV2} \citep{InceptionResNetV2}
\item \texttt{MobileNet} \citep{MobileNet}
\item \texttt{MobileNetV2} \citep{MobileNetV2}
\end{itemize}
Fig.~\ref{fig:arch} shows a sample architecture using \texttt{VGG16} for both frequency-time and DM-time models. All the above models expect three colour-channel (i.e.~RGB) images. In order to make our input data compatible with these models, we apply three $(2 \times 2)$ convolutional filters with a Rectified Liner Unit (ReLU) activation function. This is denoted as FT conv0 and DM conv0 in Fig.~\ref{fig:arch}, where FT corresponds to frequency-time, and DMT to DM-time. Note that both the FT and DMT images were scaled to zero median and unit standard deviation. The output is then attached to the above-stated models, and the top classification layer is replaced with a dense layer with two units and a softmax activation function. The softmax function takes an $N$-dimensional vector with elements $a_j$ as the input. The corresponding element-wise operation  
\begin{equation}
    S_j = \frac{e^{a_j}}{\sum_{k=1}^{\mathrm{N}}{a_j}}\, \forall j \in 1\dots \mathrm{N}
\end{equation}
makes sure that the output probabilities always sum to unity.

\begin{figure*}
	\includegraphics[width=170mm]{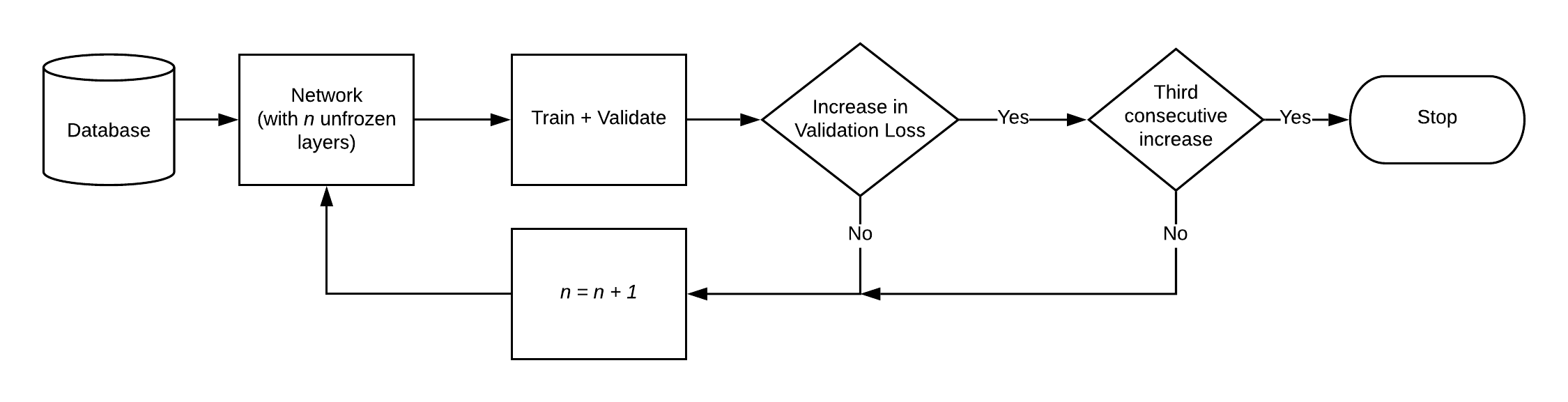}
    \caption{Flow chart explaining the training procedure followed for each frequency-time and DM-time model (see \S\ref{sec:training} for details). Here, $n$ corresponds to the number of unfrozen layers in the model, and the database corresponds to the training and validation data. We begin the training process at $n=0$, and its validation loss is recorded. We then increment $n$, i.e. unfreeze a layer and train the model. This process continues until the validation loss increases for three consecutive values of $n$. The model with the least validation loss is chosen for subsequent use.}
    \label{fig:fetch_training}
\end{figure*}

\subsubsection{Training}
\label{sec:training}
For training, we use transfer learning in the following manner. The networks with Imagenet weights are frozen, and the rest of the weights are trained and validated. The frozen weights are not modified during backward propagation. This is done because the trained models are already good at feature extraction. The training continues until the validation loss stops decreasing for at least three consecutive epochs. At this point, the model is considered to be trained. In order to tune our models further, we start unfreezing the top layers one by one and repeat the above procedure to train the network. We denote $n$ as the number of layers unfrozen. The unfreeze--train process continues till the validation loss stops decreasing for at least three trainable layers and the model configuration with least validation loss is selected (see Figure~\ref{fig:fetch_training}). To prevent the network from learning undesirable background features and overfitting, we add Gaussian noise with zero mean and unit standard deviation to the input data at each epoch. See \citet{white_noise} for a detailed analysis and discussion of the addition of white noise while training. The whole procedure is repeated separately for frequency-time and DM-time inputs.

For training, we use the Adaptive Moments (Adam) optimiser \citep{kingma2014} with a binary cross-entropy cost function. The learning rate for Adam is set to be the same as for the Imagenet training. The data are split randomly into train and validate sets, which encompass 85 and 15\% of the data, respectively. 

\begin{figure*}
	\includegraphics[width=170mm]{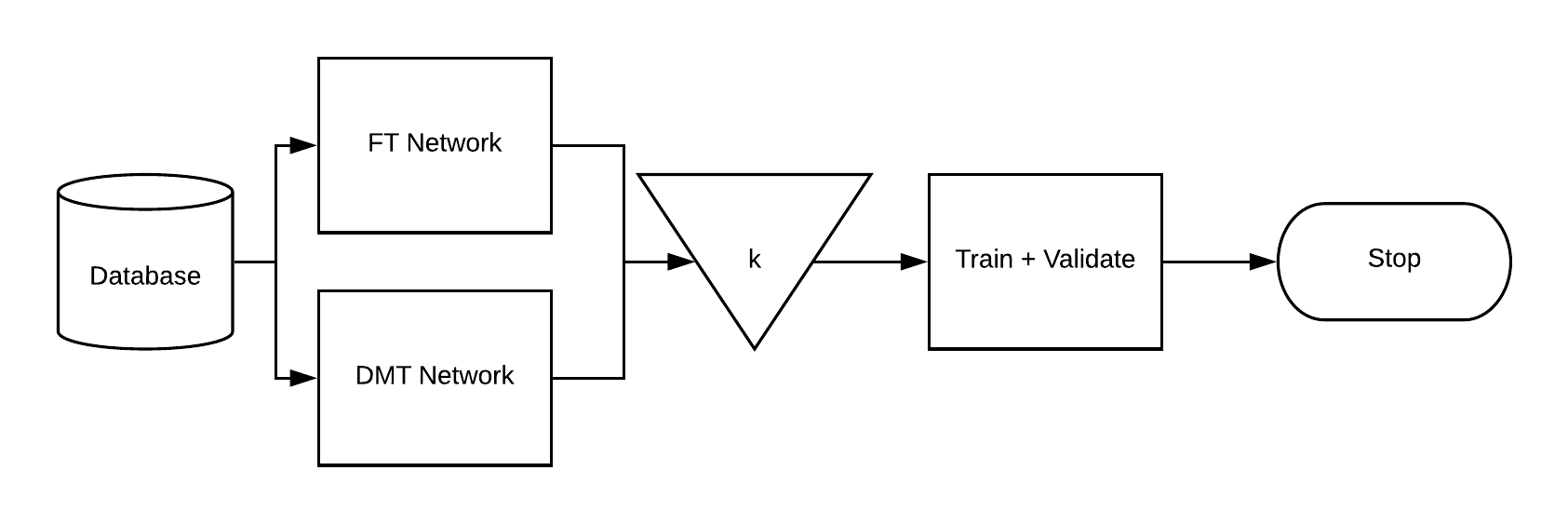}
    \caption{Flow chart explaining the training procedure followed after frequency-time and DM-time models were fused. Here $k$ corresponds to the fusion parameter. As explained in \S\ref{sec:fusion}, each pair of FT and DMT network, was fused using a dense layer with $k$ units. The database corresponds to the training and validation data.}
    \label{fig:fetch_whole}
\end{figure*}

\subsubsection{Network Fusion}
\label{sec:fusion}
Once both DM-time and frequency-time models are trained, we must combine them to get a more robust network for FRB--RFI classification. Network combination can be performed in many ways. The most common approach is to concatenate the feature extraction layer and add a classification layer. However, the layer concatenation approach did not work for us, as it over-fitted our data. 

Instead, we use the multiplicative fusion approach to fuse the two networks (see \citet{Park_fusion} and references therein). For each DM-time and frequency-time model, the top classification layer is removed. A new dense layer with $k$ units is attached to both the models. An element-wise product is then taken, followed by a classification layer with two units with a softmax activation function (softmax is described above in \S\ref{sec:network_arch}). This method allows us to combine both models with a single hyperparameter $k$, and also acts as a regularizer while training. We keep the previously trained layers unfrozen, and both the models learn simultaneously while training (see Figure~\ref{fig:fetch_whole}).

As an example, when we combined our top models for both DM-time and frequency-time, \texttt{VGG16} and \texttt{VGG19}, respectively using concatenation, the combined model yielded training accuracy of 99.11\% while the validation accuracy was 77.61\%. This is a classic case of overfitting. In contrast, combining these models using multiplicative fusion with $k=128$ lead to a training and validation accuracy of 99.9\% and 99.8\% respectively.

The training procedure detailed above is executed for all model combinations for five values of $k=(2^5,2^6,2^7,2^8,2^9)$. Based on the performance of the models with the above $k$ values, some intermediate values of $k$ were also used in some cases.
We trained our models on a Tesla P100 GPU at the XSEDE Pittsburgh Supercomputing Center. Training frequency-time and DM-time models for $\sim$10 epochs usually completed within 1~h. Training the fused networks for $\sim$10 epochs took about 1.5~h.

\subsection{Metrics}
\label{subsec:metrics}

Various metrics could be employed for evaluating the performance of the models. Our primary goal is to have these algorithms accurately identify FRBs while minimising the presentation of RFI as a good FRB candidate. We have used accuracy, precision, recall, and fscore to eliminate models, and decide what models rank highly in this regard. \textit{Accuracy} is the ratio of the number of correct predictions (of FRBs and RFIs) to the total number of predictions. \textit{Precision} is the number of FRBs correctly labelled divided by all the candidates labelled as FRBs. \textit{Recall} is the fraction of FRBs correctly classified as FRBs. \textit{Fscore} is the harmonic mean of precision and recall and is usually used to find a balance between the two. Single pulses from FRBs and pulsars are considered ``real" or ``positive" while RFI is considered ``bogus" or ``negative" for the calculation of metrics.
All metrics were computed for training and validation dataset corresponding to each model iteration. This was also used to eliminate models which suffered from overfitting (e.g.~ResNet) and underfitting (e.g.~MobileNets). 

\section{Results}
\begin{table}
    \caption{Top-5 models for frequency-time (top) and DM-time (bottom) with their respective validation accuracies (Val Acc). Number of unfrozen layers ($n$) is written in parenthesis for each model}
    \centering
    \label{tab:top-5}
        \begin{tabular}{lc}
    		\hline
    		FT Model & Val Acc (\%) \\
    		\hline
    	    \texttt{VGG19} (4) & 99.78\\
    		\texttt{VGG16} (4) & 99.40\\
    		\texttt{DenseNet169} (11) & 95.40\\
    		\texttt{DenseNet201} (7) & 94.05\\
    		\texttt{DenseNet121} (4) & 88.23\\
    		\hline        
    		\hline
    		DMT Model & Val Acc (\%) \\
    		\hline
    		\texttt{VGG16} (2) & 99.92 \\
    		\texttt{Xception} (21) & 99.87\\
    		\texttt{VGG19} (0) & 99.73 \\
    		\texttt{InceptionV3} (31) & 99.46\\
    		\texttt{InceptionResNetV2} (34) & 99.35\\
    		\hline
        \end{tabular}
\end{table}
\subsection{Model selection}
As mentioned in the previous sections, we trained many different models (see \S\ref{sec:network_arch}) individually on DM-time images, and frequency-time images. For each model, a hyperparameter $n$ (i.e., the number of trainable layers) was also found. We used \emph{validation accuracy} to decide the top-five models each for the two inputs, as this metric fulfills the most fundamental requirement: that as few as possible candidates are wrongfully classified. The metrics for these five models are given in Table~\ref{tab:top-5}. 

Twenty-five pairs of models were formed using the top-five models selected for each input. Each such pair was combined using five different values of hyper-parameter $k$, as explained in \S \ref{sec:fusion}. Additional $k$ values between the given range were also used in some cases, if the model combination was observed to perform well. Models were then filtered by their test metrics i.e accuracy, recall and fscore $>$ 99.5\%. Of the model combinations with different $k$ values, only the one with highest fscore was retained. 11 models passed this filter criterion, and are henceforth referred to as top-11 models. These top-11 models are given in Table~\ref{tab:top-11_combo}. Models in Table~\ref{tab:top-11_combo} have been sorted by the accuracy on test data. As can be observed, model \texttt{a} is the best performing model with accuracy, recall and fscore $\sim$99.9\%.

\subsubsection{Two Phase Training Approach}
In this analysis, we have opted for a two-phase training approach. The first phase involved training the frequency-time and DM-time models separately. In the second phase, we combine the models using multiplicative fusion and train them. This approach was taken to reduce the number of models to be trained.

In the case of a single-phase approach, one would start with 11 model architectures for both frequency-time and DM-time models. Combining them with five values of fusion hyperparameter $k$, would result in 605 combined models. Then for each model, one would unfreeze $n_1$ and $n_2$ number of layers in frequency-time and DM-time model respectively. If on an average, 10 layers were unfrozen for both the models (i.e. $n_1 = n_2 = 10$), it would result in 60,500 models, each of which would have to be trained separately.

In contrast, the two-phase approach would consist of training 110 models for both frequency-time and DM-time each in the first phase. In the second phase, combining top-5 frequency-time and DM-time models with five values of hyperparameter $k$ would lead to 125 models to be trained. As a result, one would train 345 models with this approach, which is much less than the number of models to be trained using single-phase approach.

\begin{table*}
	\centering
	\caption{Top-11 models with their corresponding metrics on test data. Again, number of unfrozen layers ($n$) is written in parenthesis for each model. $k$ is the fusion hyperparameter. FT, DMT corresponds to frequency-time and DM-time.}
	\label{tab:top-11_combo}
	\begin{tabular}{ccccccc} 
		\hline
		Label & FT Model & DMT Model & $k$ & Accuracy (\%) & Recall (\%) & Fscore (\%)\\
		\hline
        a & \texttt{DenseNet121} (4) & \texttt{Xception} (21) & 256 & 99.88 & 99.92 & 99.87\\
        \hline
        b & \texttt{DenseNet121} (4) & \texttt{VGG16} (2) & 32 & 99.86 & 99.92 & 99.85\\
        \hline
        c & \texttt{DenseNet169} (11) & \texttt{Xception} (21) & 112 & 99.86 & 99.78 & 99.85\\
        \hline
        d & \texttt{DenseNet201} (7) & \texttt{Xception} (21) & 32 & 99.86 & 99.78 & 99.85\\
        \hline
        e & \texttt{VGG19} (4) & \texttt{Xception} (21) & 128 & 99.85 & 99.75 & 99.84\\
        \hline
        f & \texttt{DenseNet169} (11) & \texttt{VGG16} (2) & 512 & 99.81 & 99.7 & 99.79\\
        \hline
        g & \texttt{VGG19} (4) & \texttt{VGG16} (2) & 128 & 99.79 & 99.59 & 99.77\\
        \hline
        h & \texttt{DenseNet201} (7) & \texttt{InceptionResNetV2} (34) & 160 & 99.76 & 99.72 & 99.74\\
        \hline
        i & \texttt{DenseNet201} (7) & \texttt{VGG16} (2) & 32 & 99.75 & 99.59 & 99.73\\
        \hline
        j & \texttt{VGG19} (4) & \texttt{InceptionResNetV2} (34) & 512 & 99.68 & 99.59 & 99.65\\
        \hline
        k & \texttt{DenseNet121} (4) & \texttt{InceptionV3} (31) & 64 & 99.66 & 99.62 & 99.63\\
        \hline
	\end{tabular}
\end{table*}
\subsection{Evaluating Performance on Independent Data (and Actual FRB Detections)}
\label{subsec:eval_real}
We evaluated the performance of our top-11 models on independent FRB data. This serves a two fold purpose. First, it would demonstrate how well our models perform on real FRBs, as they were trained on pulsars and simulated FRBs. Second, this would show how well the models would generalise to data from other telescopes. Given that each telescope has its unique instrumental effects and RFI environment, it is imperative to do such tests to gain confidence in the performance of the models in potentially vastly different RFI environments. This can be considered as a real world test dataset as it is representative of typical FRB-searches.

\subsubsection{Data}
We used the FRB data from ASKAP \citep{shannon18}, Parkes (5 from \citep{champion2016}, FRB~110220 \citep{thronton13}, FRB~150215 \citep{petroff2017} and FRB~140514 \citep{petroff2015}) and FRB~121102 data from Breakthrough Listen \citep{gajjar2018, zhang18}. We used only 8 out of 22 Parkes FRBs, as the rest of them had 96 frequency channels. These datasets were fed to the transient detection pipeline, {\sc heimdall}, which uses sliding boxcar filters to search for transients at various widths and S/N thresholds and is in standard use in multiple FRB search pipelines  around the world. Candidates which meet the following search criterion were produced: $\text{S/N} \geq 8, 10<\rm{DM}<10000~\dmunits, \text{width} < 32$~ms. The candidates thus produced were inspected visually. 

Out of the 10,672 candidates found from ASKAP data, we selected the 33 FRB detections (20 unique FRBs, a few detected in multiple beams) reported in \citet{shannon18}. The remainder of the 10,639 candidates were manually parsed through for verification and labelled as RFI. From Parkes data, we obtained 486 candidates (8 we marked as FRBs, 478 as RFI). From Breakthrough Listen data, we obtained 15 pulses of FRB~121102, and the remaining 652 candidates were labelled as RFI. 

\subsubsection{Model performance}
In table~\ref{tab:top-11_eval}, we report  the number of correct classifications of FRBs and incorrect classifications of RFI.  All of the models were able to classify all the ASKAP and Parkes FRBs except model \texttt{b,g}. While for FRB~121102, four models were able to classify all the pulses correctly. Moreover, the rate of mislabelling RFI as FRB was relatively low, as evident in the table. The precision and recall values for the same are reported in table~\ref{tab:top-11_pr_vals}.

Note that these models were not trained on data from any of these back-ends, which is a testament to the instrument-agnostic capabilities of our trained algorithm, which appears to be relatively transferable despite the lack of re-training. Performance can be further improved by training the models with a few thousand candidates from any new back-end. This procedure is detailed in \S\ref{subsec:fine_tune}.

The satisfactory performance of our models on data from these different back-ends provides reasonable confidence that they have learned features about RFI and FRBs that are sufficiently general such that they can distinguish an FRB from RFI, using only the frequency-time and DM-time images.


\begin{table*}
	\centering
	\caption{Results of model evaluation on Real FRB data from ASKAP, Parkes and Breakthrough Listen (BL) backend. Total number of candidates in each case is written alongside the title. Numbers in the bold represent the best performing models for the corresponding cases.}
	\label{tab:top-11_eval}
	\begin{tabular}{ccccccc} 
		\hline
		Label & ASKAP FRBs  & Mislabelled AKSAP & Parkes FRBs  & Mislabelled Parkes & BL 121102  & Mislabelled BL \\
		& (/33) &  RFI (/10639) & (/8) & RFI (/478) & (/15) & RFI (/652)\\
		\hline
        a & \textbf{33} & 2 & \textbf{8} & \textbf{0} & 14 & \textbf{0}\\
        \hline
        b & 28 & 5 & \textbf{8} & 48 & \textbf{15} & 1\\
        \hline
        c & \textbf{33} & 16 & \textbf{8} & 6 & \textbf{15} & \textbf{0}\\
        \hline
        d & \textbf{33} & 12 & \textbf{8} & 29 & \textbf{15} & \textbf{0}\\
        \hline
        e & \textbf{33} & 16 & \textbf{8} & 7 & 14 & \textbf{0}\\
        \hline
        f & \textbf{33} & 2 & \textbf{8} & 1 & 14 & \textbf{0}\\
        \hline
        g & 29 & \textbf{1} & \textbf{8} & 10 & 9 & 15\\
        \hline
        h & \textbf{33} & 43 & \textbf{8} & 40 & 14 & 5\\
        \hline
        i & \textbf{33} & 15 & \textbf{8} & 52 & 14 & 1\\
        \hline
        j & \textbf{33} & 33 & \textbf{8} & 3 & 13 & 45\\
        \hline
        k & \textbf{33} & 7 & \textbf{8} & 70 & \textbf{15} & 1\\
        \hline
	\end{tabular}
\end{table*}

\section{Discussion}
\subsection{Inference speeds and size}
\begin{figure}
	\includegraphics[width=85mm]{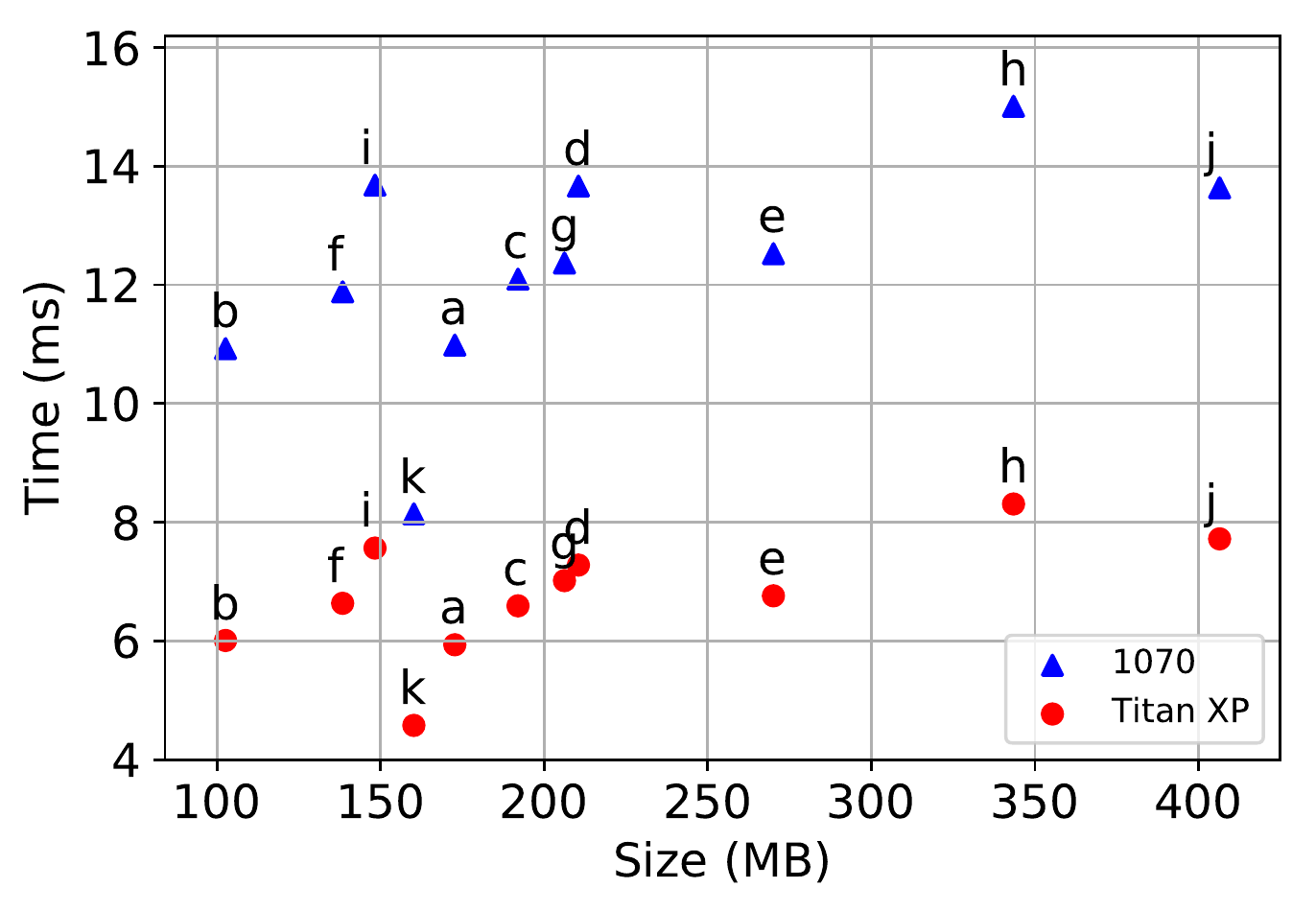}
    \caption{Time taken for classifying one candidate (in ms) with respect to the size of the model (in MB). Blue triangles represent evaluation times on NVIDIA GTX--1070, while red circles are for NVIDIA Titan--Xp. Labels \texttt{a} through \texttt{k} correspond to the models defined in Table~\ref{tab:top-11_combo}}
    \label{fig:model_speed}
\end{figure}
We measure the inference speed of our models on NVIDIA GTX--1070 and NVIDIA Titan--Xp using our test data set with a batch size of 64. For both of the GPUs, the mean times were $12 \pm 1$~ms and $6.7 \pm 0.9$~ms respectively (see Fig.~\ref{fig:model_speed}). Therefore, for a conservative time of $\sim 20$~ms per candidate, all of our top-11 models can work in real time if the candidate rate does not exceed $\sim 10^8$ per hour. Most GPU accelerated pipelines use clustering algorithms to cluster candidates in a multi--dimensional parameter space (e.g., DM, box-car width, arrival time). As a result, the number of candidates per hour is significantly smaller. As an example, using {\sc heimdall} on the $\sim$700~hours of full scan ASKAP data from \citet{shannon18}, we obtained $\sim 10^4$ candidates. Therefore any of our top-11 models could be used in a commensal pipeline for real-time classification of the candidates and triggers for multi-frequency follow-ups. However, it should be noted that ASKAP is in a radio-quiet zone. Therefore the number of RFI candidates would be smaller. 

Fig.~\ref{fig:model_speed} can also be used to compare the sizes of individual models. The size of a model is proportional to the number of parameters in the model. Hence larger models tend to run slower. While the above is generally true, it should be noted that the model architecture itself plays an essential role in the inference speed.

\begin{figure*}
	\includegraphics[width=170mm]{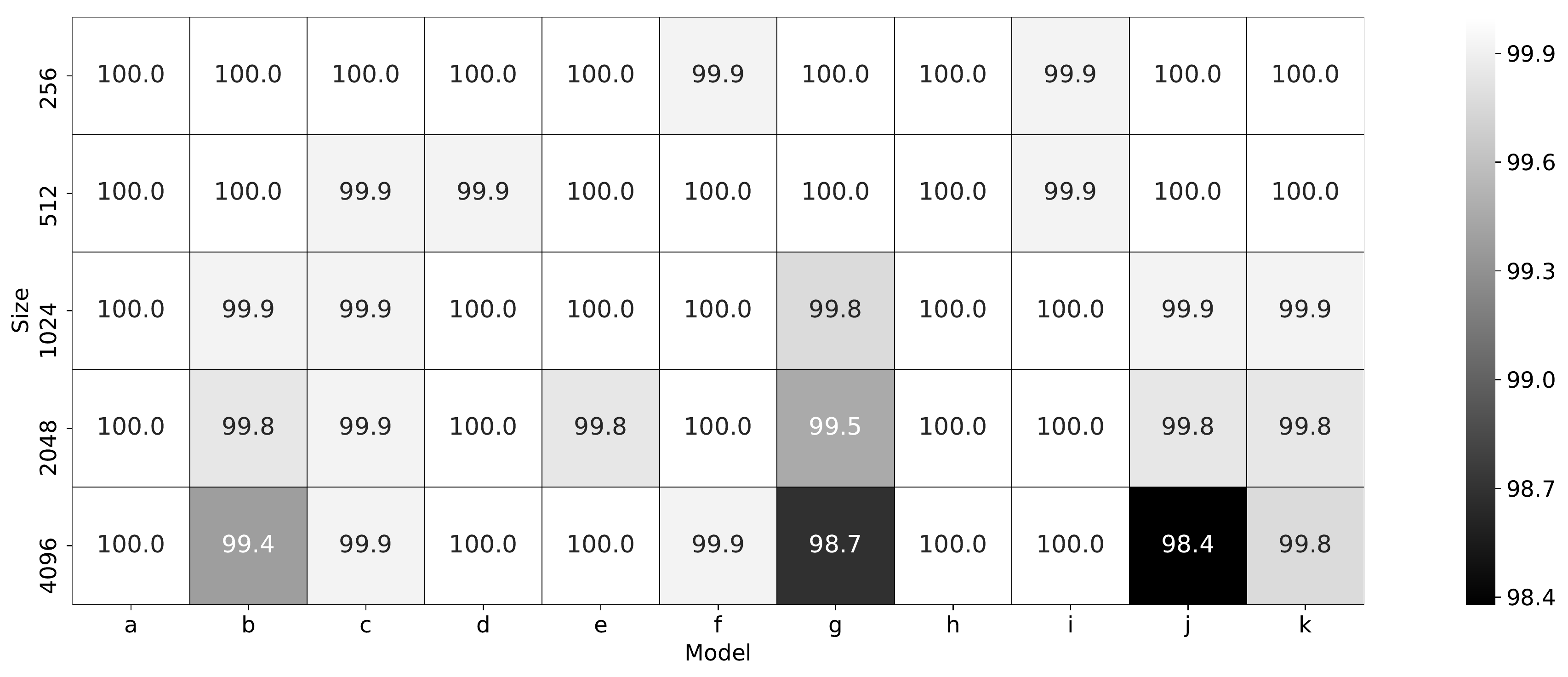}
    \caption{Heatmap for accuracies of differently sized frequency-time inputs. The accuracies are colour-coded and annotated. The time axis was kept to be 256 pixels. The Y-axis shows the number of pixels in the frequency axis. Labels \texttt{a} through \texttt{k} on the X-axis correspond to the models defined in Table~\ref{tab:top-11_combo}}
    \label{fig:snracc}
\end{figure*}
\subsection{Input shapes}
For training as well as testing, we have used 256$\times$256 pixel images for both Frequency-Time and DM-Time. As explained in \S\ref{sec:input_data_std}, to achieve that size, we applied a standardisation procedure to both images. In order to test our models for various input sizes, we used high S/N pulsar candidates from GREENBURST and binned the frequency axis to different sizes (4096, 2048, 1024, 512). We also added Gaussian noise to the data to artificially reduce its S/N, such that for each size we have a uniform distribution of S/N between 8 and 40 with $\sim$650 candidates. We also used the same number of RFI candidates for each input size. However, Gaussian noise was not added to the RFI images. We then used our top-11 models to evaluate these candidates. The results are presented as a heatmap in Fig.~\ref{fig:snracc}. This demonstrates that our models are not very sensitive to changes in image size, and only show a marginal decrease in accuracy, while the recall stayed at 100\%. As mentioned in \S\ref{subsec:DLinTD}, a larger image size could thus be used with our models to preserve the frequency modulation of FRBs. Hence, data from commensal FRB search back-ends, for example, CRAFT-ASKAP, GBTrans, UTMOST with 336, 512 and 320 frequency channels respectively, can directly be fed into the models.



\subsection{Sensitivity analysis}
\begin{figure*}
	\includegraphics[width=170mm]{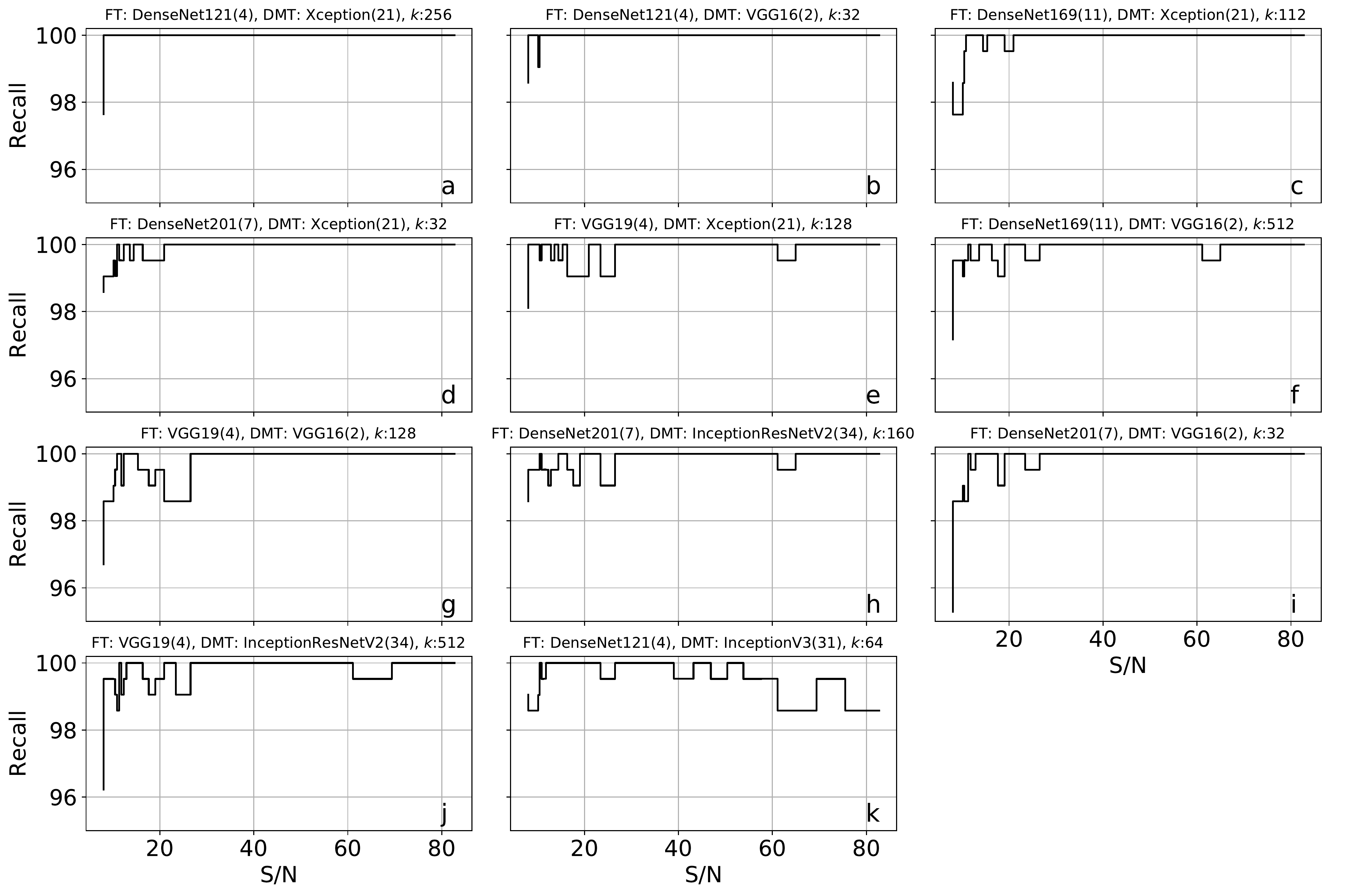}
    \caption{Recall vs Signal to noise (S/N) for top-11 models, evaluated on the test dataset. The FRBs from the dataset were binned into 30 S/N bins, each with an equal number of candidates.  Labels \texttt{a} through \texttt{k} also correspond to the models defined in Table \ref{tab:top-11_combo}}
    \label{fig:snrhistogram}
\end{figure*}

It is imperative to analyse the sensitivity of the models with respect to the S/N of the candidates. Although, the performance reported in Table~\ref{tab:top-11_combo} is useful to compare models, it is a cumulative number, i.e.~how well the models performed on the complete test data. Figure \ref{fig:snrhistogram} shows the recall as a function of S/N of the FRBs in the test dataset. To compute this, we used all the FRB candidates from the test dataset and binned them into 30 bins, each with an equal number of candidates. The top 11 models were used to classify these candidates, and recall per bin was calculated (refer to \S\ref{subsec:metrics} for details on recall calculation). As expected, recall improves as the S/N increases, as it is easier to classify higher S/N candidates. For most of our cases, the recall remained $>99\%$ above a S/N of 10 (except model \texttt{g} and \texttt{k}). We also note that, due to the limited amount of data, each bin only had a few hundred candidates, which are statistically not enough to quantify such a trend. Hence these recall values per bin should be taken with caution, and the figure should only be interpreted qualitatively. Typically, we would like to have several thousand candidates per bin in order to produce robust and reliable metrics.

\subsection{Fine tuning}
\label{subsec:fine_tune}
While our models perform well on data from different telescopes and backends, it is still possible to further improve their performance for a specific use case. The models can be fine-tuned by re-training their final classification layer using few thousand candidates. In order to demonstrate this, we decided to use the data recorded at a frequency other than L-band, as all our models were originally trained on L-band data. For this purpose, we used the observations of FRB121102 recorded using Breakthrough Listen Digital Backend at 4--8~GHz \citep{gajjar2018}.

We re-purpose the 652 RFI candidates as mentioned in \S \ref{subsec:eval_real}. Using the procedure described in  \S\ref{subsec:simulated_frb} we generated 700 simulated FRB candidates at 4--8~GHz with the above-specified data as the background. 80\% of this data was used for training, and 20\% was marked for validation. The final classification layer was trained using the procedure described in \S\ref{sec:training}. To compare the performance of the fine-tuned models, we re-evaluate them on the 15 FRB~121102 pulses as shown in table \ref{tab:top-11_eval}. After fine tuning, all of our models (except model \texttt{g}) were able to correctly classify at least 14 out of 15 pulses, with six models classifying all 15 pulses correctly. This whole exercise took $\sim 15$~min per model on an NVIDIA GTX--1070Ti GPU.



\subsection{Comparison to previous work}


In order to compare different machine learning algorithms in a fair manner, they should be evaluated on a common standard data set. As only a handful of FRBs has been detected to date, such a dataset cannot be created with real data. This has been discussed in great detail by \citet{connor18}. Also, machine learning algorithms like Support Vector Machines \citep{hearst1998} and Random Forest \citep{breiman2001} take advantage of the features, which are custom made to the specific telescope or survey. For example, the antenna covariance and network dropouts in the V-FASTR algorithm or relative candidate MJD information in the ALFABURST algorithm. Excluding such features would lead to performance degradation of the respective classifier, thereby rendering the final comparison inconclusive. Realising the need for a standardised dataset, we provide our dataset\footnote{\url{http://astro.phys.wvu.edu/fetch/}} for testing future algorithms.

For the sake of completeness, we present a weak comparison between the \citeauthor{connor18} network by training and testing it on our data. We emphasise the fact that the authors trained their network on CHIME and LOFAR data independently, whereas our dataset contains a mixture of backends. We use the data as reported in table \ref{tab:datasets} and resize the images to (32, 64) pixels for frequency-time and (64, 64) for the DM-time. We omit the multi-beam S/N and pulse profile part of their network and train the merged model following the same procedure as reported by the authors. Pulse profile input wasn't included as it did not improve the test accuracy. Evaluating their model on the test data as reported in table \ref{tab:datasets}, the accuracy, recall and fscore were 97.96\%, 95.76\% and 97.81\% respectively. When compared on a common data set, our models show better performance. The differences in the performance elucidate two key features of our study -- the importance of deeper neural networks and transfer learning. Transfer learning enabled the use of state of the art neural networks for our application. These deep networks, extracted more generalised features and thus proved better at classification.
\subsection{\texttt{FETCH}}
\label{subsec:fetch}
We provide a user-friendly open-source python package \texttt{FETCH} (Fast Extragalactic Transient Candidate Hunter)\footnote{\url{https://github.com/devanshkv/fetch}}, for real-time classification of candidates from single pulse search pipelines, using our top-11 models. The input of \texttt{FETCH} is a candidate file containing the frequency-time and DM-time data. For each candidate and a choice of model, it outputs the probability of the candidate to be an FRB. These candidate files can be generated from filterbanks using \texttt{pysigproc}\footnote{\url{https://github.com/devanshkv/pysigproc}}. 

Using \texttt{FETCH}, the classification probabilities from all 11 models can be combined using simple mathematical operations like averaging, intersection, union or majority voting. This would result in a more robust classification. This approach however is slower, and requires more computational resources. If only one model has to be used, the model \texttt{a} should be chosen, as it performs the best on our metric (see Table~\ref{tab:top-11_combo}).
\texttt{FETCH} also provides a framework to fine-tune the models to further improve its performance for particular backends. As demonstrated in \S\ref{subsec:fine_tune}, this can be done with a few thousand labelled candidates. It is recommended to use a balanced dataset, wherein the number of RFI and FRB candidates are comparable. 

Presently, \texttt{FETCH} is being integrated into the GREENBURST pipeline and {\sc realfast} for commensal FRB searches at the GBT and Very Large Array telescope respectively. For  {\sc realfast}, along with frequency-time and DM-time networks and  \texttt{FETCH} will feature an additional third network with radio image as an input.
\subsection{Future work}
Here, we discuss a few potential techniques for improvement of our models, which would be pursued in future.
\subsubsection{Model pruning}
\begin{figure}
    \centering
	\includegraphics[width=\columnwidth]{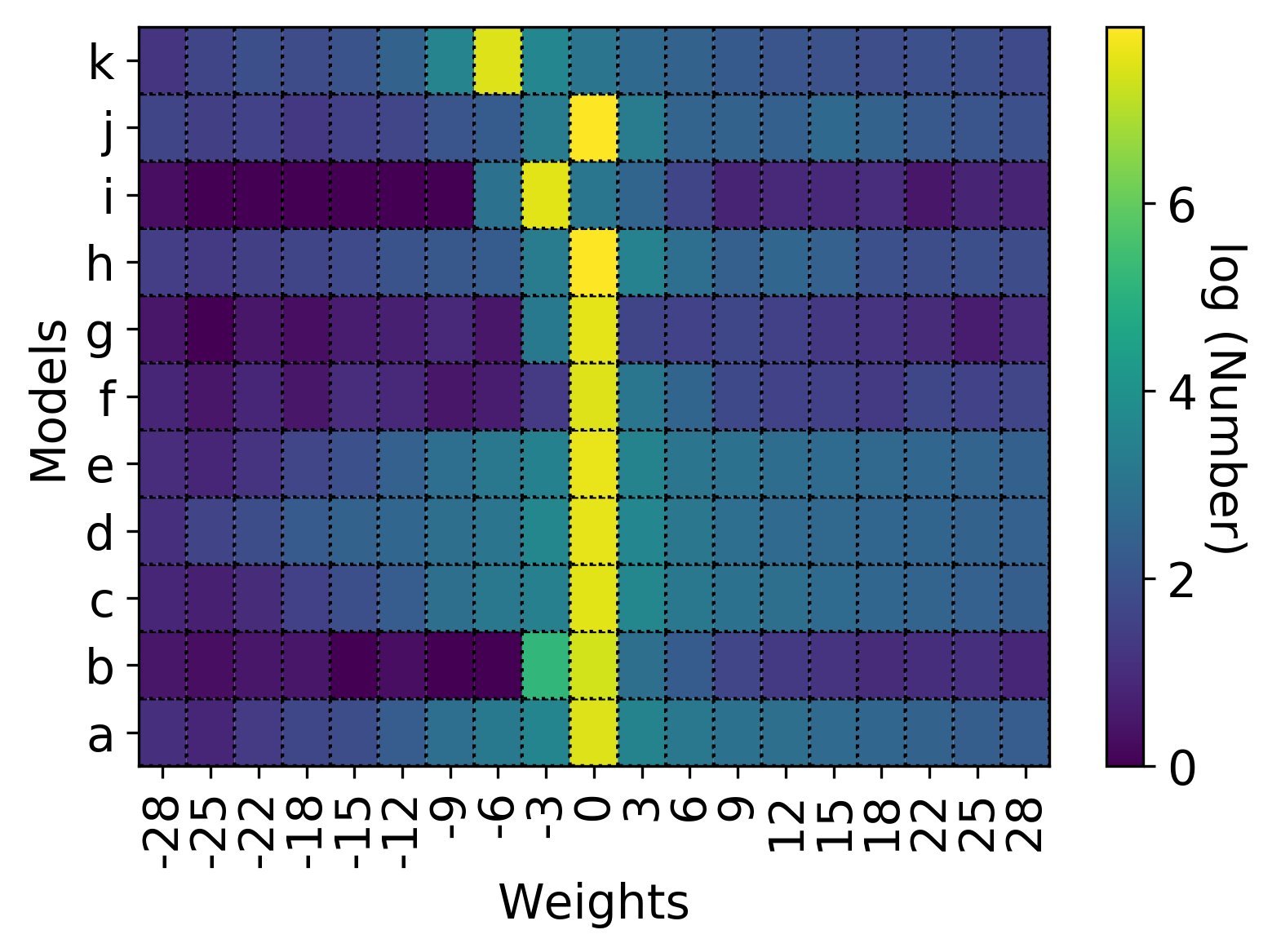}
    \caption{Distribution of weights for all the models. The x-axis shows the model weights while the models are plotted on the y axis. The colorbar denotes the log of number for each model. Notice that most of the weights are nearly zero, and can be pruned, reducing the size of model significantly.}
    \label{fig:model_weights}
\end{figure}
In Fig.~\ref{fig:model_weights}, we show the distribution of weights from all the layers, for each model. As is evident from these distributions, a large number of weights are near zero. The network connections with such weights can be pruned, and models can be retrained. This technique has shown that models like \texttt{VGG16/19} can be compressed up to 10 times, without loss in performance (see \citet{Han_Pruning} for details).

\subsubsection{Model quantization}
Presently, all of our model weights are 32-bit floats. The network weights can be quantised to 16-bit floats or 8-bit integers and retrained, leading to faster and smaller sized networks. Typically this offers twice the speed up without performance loss. We suggest the reader to see \citet{Han_deep_compression} for a more detailed description.

\subsubsection{Adversarial noise}
Machine Learning algorithms often misclassify data when presented with adversarial examples. That is, the algorithm fails when a ``carefully computed'' adversarial noise term is added to a previously correctly classified example. The new perturbed input can be written as,
\begin{equation}
    \Hat{\bold{x}} = \bold{x} + \delta \,\rm{sgn}( \nabla_x J(\bold{x})).
\end{equation}
Here $\bold{x}$ is the original input, and $J(\bold{x})$ is the cost function, $\mathrm{sgn}$ is the signum function and $\delta$ is a very small number (e.g.~for 8-bit integer input data, $\delta \sim 0.1$ is used). Such small perturbations are indistinguishable to the human eye. See \citet{Goodfellow_Adv} for more detailed analysis of adversarial noise. In order to make our models more robust, we can also add adversarial noise while training.

\section{Conclusions}
We have presented 11 deep learning models to classify FRB and RFI candidates. Using the technique of transfer learning, we trained state-of-the-art models on frequency-time and $\rm DM$--time images individually. These models were then combined using multiplicative fusion in order to improve performance. We have used L-Band data from the GBT and 20~m telescope at the GBO to train our models. All models perform with accuracy and recall $>$99.5\% on our test dataset. These models are frequency and telescope agnostic, and the majority of them detected all the FRBs from ASKAP and Parkes telescope and FRB121102 pulses above an S/N of 8. We also show that the models can be fine-tuned to a specific backend by re-training them with $\sim$1000 labelled examples, to improve their performance further. 

We provide a python based open source package \texttt{FETCH} for the classification of candidates using our models. The average classification time of our models is $12 \pm 1$~ms per candidate on NVIDIA GTX--1070Ti. Therefore using \texttt{FETCH} our models can be promptly deployed at any commensal FRB search backends and can be used to send real-time triggers for multi-frequency follow up.

\section*{Acknowledgements}
D.A.\ and K.A.\ would like to thank Shalabh Singh for useful discussions. The authors also thank Vishal Gajjar for providing help with processing the Breakthrough Listen data. The authors also thank Casey Law for comments on the manuscript. K.A.\ and S.B.S.\ are supported through the National Science Foundation (NSF) award AAG-1714897. D.R.L. also acknowledges support from the Research Corporation for Scientific Advancement and NSF award AAG-1616042. All authors acknowledge support from the NSF awards OIA-1458952 and PHY-1430284. 

The authors acknowledge use of the Super Computing System (Spruce Knob) at the WVU, which is funded in part by the National Science Foundation EPSCoR Research Infrastructure Improvement Cooperative Agreement \#1003907, the state of West Virginia (WVEPSCoR via the Higher Education Policy Commission) and WVU. This work used the Extreme Science and Engineering Discovery Environment (XSEDE) \citep{towns2014}, which is supported by National Science Foundation grant number ACI-1548562. Specifically, it used the Bridges system \citet{nystrom2015}, which is supported by NSF award number ACI-1445606, at the Pittsburgh Supercomputing Center. Finally, the authors thank the anonymous referee for helpful comments.

\section*{Data availability}
The data and software underlying this article are available in Zenodo, at \url{https://dx.doi.org/10.5281/zenodo.3905437}




\bibliographystyle{mnras}
\bibliography{frbml}



\appendix

\section{Precision and recall values on the real data}
In addition to table \ref{tab:top-11_eval} we provide the precision and recall values based on the metrics discussed in \S\ref{subsec:metrics}. While these metrics are from the actual FRB search data, these numbers should be interpreted cautiously because of the small number statistics.
\begin{table}
	\centering
	\caption{Precision and recall values on real FRB data from ASKAP, Parkes and Breakthrough Listen (BL) backend. Here R and P correspond to recall and precision respectively.}
	\label{tab:top-11_pr_vals}
	\begin{tabular}{ccccccc} 
		\hline
	    \multirow{2}{*}{Label} &
	    \multicolumn{2}{c}{ASKAP} &
	    \multicolumn{2}{c}{Parkes} &
	    \multicolumn{2}{c|}{BL 121102} \\
		& R & P & R & P & R & P\\
		\hline
        a & 1.00 & 0.94 & 1.00 & 1.00 & 0.93 & 1.00 \\
        \hline
        b & 0.85 & 0.85 & 1.00 & 0.14 & 1.00 & 0.94 \\
        \hline
        c & 1.00 & 0.67 & 1.00 & 0.57 & 1.00 & 1.00 \\
        \hline
        d & 1.00 & 0.73 & 1.00 & 0.22 & 1.00 & 1.00 \\
        \hline
        e & 1.00 & 0.67 & 1.00 & 0.53 & 0.93 & 1.00 \\
        \hline
        f & 1.00 & 0.94 & 1.00 & 0.89 & 0.93 & 1.00 \\
        \hline
        g & 0.88 & 0.97 & 1.00 & 0.44 & 0.60 & 0.38 \\
        \hline
        h & 1.00 & 0.43 & 1.00 & 0.17 & 0.93 & 0.74 \\
        \hline
        i & 1.00 & 0.69 & 1.00 & 0.13 & 0.93 & 0.93 \\
        \hline
        j & 1.00 & 0.50 & 1.00 & 0.73 & 0.87 & 0.22 \\
        \hline
        k & 1.00 & 0.82 & 1.00 & 0.10 & 1.00 & 0.94 \\
        \hline	
    \end{tabular}
\end{table}

\bsp	
\label{lastpage}
\end{document}